\documentstyle[11pt]{article}
\begin{document}

\title{Photometric Parameters of RR~Lyr Variable Stars in the Galactic Bulge}
\author{Arkadiusz ~~ O~l~e~c~h}

\date{Warsaw University Observatory, Al.~Ujazdowskie~4,~00-478~Warszawa,
Poland\\
e-mail:olech@sirius.astrouw.edu.pl}
\maketitle 
 
\abstract{We present a complete set of OGLE photometric measurements for 214 
RR~Lyrae variables in the Galactic bulge, based on four observing seasons: 
1992--95. These are mostly $I$-band measurements, with some 
$V$-band observations obtained mainly in 1995. Based on this material we 
construct the Fourier decomposition of the $I$-band light curves and 
investigate different types of RR~Lyr variability. We also determine the value 
of the extinction-insensitive parameter $V_{V-I}$ for RR~Lyr stars in fields 
with different galactic longitude $l$ and compare with values of $V_{V-I}$ 
calculated for red clump stars by Stanek {\it et al.} (1994). 

The observed and free from interstellar extinction mean brightness and ${V-I}$ 
color of RR~Lyr stars are also presented.}


\section{Introduction}

The Optical Gravitational Lensing Experiment (OGLE) is a long term observing 
project which main goal is an extensive photometric search for gravitational 
microlensing events (Paczy\'nski 1986, Udalski {\it et al.} 1992). For this purpose 
the CCD photometry of a few millions stars in dense regions near the center of 
Galaxy was performed. Up to now, after four seasons of observations (years 
1992--1995) nineteen microlensing events have been detected (Udalski {\it et al.} 
1993b, 1994a, 1994b, Paczy\'nski and Udalski 1996). All the data were obtained 
with the 1-m Swope telescope at the Las Campanas Observatory in Chile, 
operated by Carnegie Institution of Washington with a ${2048\times2048}$ 
Ford/Loral CCD detector with scale 0.44"/pixel. 

Such a large amount of precise photometric data give us an opportunity to 
study different types of stellar variability (Udalski {\it et al.} 1994c, 1995a, 
1995b, 1996, 1997), to obtain good quality color magnitude diagrams (Udalski 
{\it et al.} 1993a), to construct a catalog of all stars observed in Baade's Window 
(Szyma\'nski {\it et al.} 1996) or to study the Galactic structure (Stanek {\it et al.} 
1994, 1997). Data collected during the search for periodic variable stars in 
the central part of our Galaxy were published as The Catalog of Periodic 
Variable Stars in The Galactic Bulge (Udalski {\it et al.} 1994c, 1995a, 1995b, 1996, 
1997). The Catalog contains $I$-band light curves, ${V-I}$ colors, periods, 
equatorial coordinates of stars with ${\langle I\rangle}$ and period in range 
14--18~mag and 0.1--100~days respectively. 

Discovered variable stars were grouped into three categories: pulsating 
stars (mostly RR~Lyr and $\delta$~Sct stars), eclipsing stars (mostly W~UMa, 
$\beta$~Lyr and Algol type) and miscellaneous variables (mostly late type, 
chromospherically active stars, ellipsoidal variables and a few Miras). 

Investigation of eclipsing and miscellaneous stars from OGLE data have been 
already performed (Ruci\'nski 1997a, 1997b, Olech 1996). The main aim of this 
paper is to provide more information about photometric parameters of RR~Lyr 
variable stars from the Galactic bulge. 

\section{RR~Lyr Stars in OGLE Catalog of Periodic Variable Stars}

\subsection{Statistics}

Recently OGLE Collaboration published five parts of their Catalog of Periodic 
Variable Stars in the Galactic Bulge (Udalski {\it et al.} 1994c, 1995a, 1995b, 1996, 
1997). Precise light curves, colors and finding charts for almost 3000 
variable stars were presented. Among  them 214 RR~Lyr stars were detected. 
Table~1 summarizes results presented in the Catalog and gives galactic

\vspace{5pt}
\begin{center}
Table 1 \\
\vspace{3pt}
 
RR Lyr Stars in OGLE Fields. \\
\vspace{7pt}
 
{\footnotesize
\begin{tabular}{|l|r|r|c|}
\hline
\hline
Field & $l$~~~ & $b$~~~ & RR~Lyr \\
\hline
\hline
BWC & $1.0^\circ$ & $-3.9^\circ$ & 27 \\
BW1 & $1.1^\circ$ & $-3.6^\circ$ & 16 \\
BW2 & $0.7^\circ$ & $-3.8^\circ$ & ~8 \\
BW3 & $0.9^\circ$ & $-4.2^\circ$ & 15 \\
BW4 & $1.3^\circ$ & $-4.0^\circ$ & 14 \\
BW5 & $0.9^\circ$ & $-3.7^\circ$ & 13 \\
BW6 & $0.8^\circ$ & $-4.0^\circ$ & 13 \\
BW7 & $1.1^\circ$ & $-4.1^\circ$ & 12 \\
BW8 & $1.2^\circ$ & $-3.8^\circ$ & 11 \\
BW9 & $0.9^\circ$ & $-3.3^\circ$ & 16 \\
BW10 & $0.7^\circ$ & $-3.4^\circ$ & 21 \\
BW11 & $0.5^\circ$ & $-3.5^\circ$ & 12 \\
\hline
MM5-A & $-4.8^\circ$ & $-3.4^\circ$ & 14 \\
MM5-B & $-4.9^\circ$ & $-3.5^\circ$ & 10 \\
\hline
MM7-A & $5.4^\circ$ & $-3.3^\circ$ & ~8 \\
MM7-B & $5.5^\circ$ & $-3.5^\circ$ & 10 \\
\hline
\hline
\end{tabular}
}
\end{center}
\vspace{5pt}

\noindent coordinates of the OGLE ${15'\times15'}$ fields and number of RR~Lyr stars 
detected in each of them. One can notice that number of RR~Lyr stars in BWC 
field is significantly larger than in other fields. This is caused by the fact 
that all BW fields overlap with neighboring fields by 1 arcmin and stars 
discovered in common regions of BWC and other fields were counted as BWC 
variables. 

Data presented in this work were collected during four observational seasons 
(1992--95). Average numbers of $I$-band and $V$-band measurements amounted to 
about 100--200 (depending on the field) and over 20 respectively. Majority of 
\hbox{$V$-band} data were obtained during the 1995 season. Complete set of above 
mentioned data is available via INTERNET in electronic form (ftp host: 
sirius.astrouw.edu.pl, directory: {\it /ogle/var\_catalog}). 

\subsection{Fourier Coefficients for RR~Lyr stars in Baade's Window}

To be able to classify various light curves of RR~Lyr variables in 
some algorithmic way  we constructed a Fourier fit in the form: 
$$I=a_0+\sum_{k=1}^{3}\lbrace a_k\cdot\sin({{2k\pi(t-t_0)}\over{P}})+
b_k\cdot\cos({{2k\pi(t-t_0)}\over{P}})\rbrace\eqno(1)$$
for each $I$-band light curve of RR~Lyr star in OGLE Catalog. Coefficients 
$a_k$ and $b_k$ were calculated by least squares method and 
${t_0=2448000.0}$~HJD. The results are presented in Tables~2--6. 
Such a decomposition of light curve can be potentially used to distinguish 
between different kinds of RR~Lyr and also to determine physical properties of 
these stars e.g. absolute magnitude and metallicity. Fig.~1 gives 
\hbox{$I$-band} light curves of six RR~Lyr stars from BW1 with solid line 
showing the fit given by the Eq.~(1). 

\begin{center}
Table 2 \\
 
Fourier coefficients for BWC and BW1 RR Lyr stars. \\
\nopagebreak
{\scriptsize
\begin{tabular}{|l|c|c|c|c|c|c|c|c|}
\hline
\hline
 Star & Period & $a_0$ & $a_1$ & $a_2$ & $a_3$ & $b_1$ & $b_2$ & $b_3$
\\
\hline
\hline
BWC V6   & 0.42765 & 15.11787 & 0.01603 & 0.04570 & 0.05702  & --0.15597  & --0.04459 & 0.02319 \\
BWC V14  & 0.44022 & 15.69728 & 0.22144  & --0.10376 & 0.06836 & 0.17845  & --0.07848 & 0.08727 \\
BWC V15  & 0.45871 & 15.71623 & 0.01104 & 0.01575  & --0.02014 & 0.20202  & --0.04353 & 0.01021 \\
BWC V17  & 0.29872 & 15.33980 & 0.01980 & 0.01788  & --0.01424 & 0.12980  & --0.00286  & --0.00528 \\
BWC V22  & 0.48968 & 15.75357 & --0.23042 & 0.05772 & 0.14641 & 0.15559 & 0.18418 & 0.02135 \\
BWC V23  & 0.45426 & 15.82007 & 0.30212 & 0.03852  & --0.12980  & --0.13698 & 0.27124  & --0.05865 \\
BWC V25  & 0.47418 & 15.71959 & --0.04423  & --0.01694 & 0.02301  & --0.21444  & --0.06483  & --0.01531 \\
BWC V26  & 0.47863 & 15.83531 & 0.12667 & 0.08377  & --0.02398  & --0.20163 & 0.04397 & 0.07948 \\
BWC V28  & 0.59478 & 15.63364 & 0.03668 & 0.06184  & --0.06175 & 0.15190  & --0.03995  & --0.02804 \\
BWC V30  & 0.57147 & 15.66380 & 0.09582 & 0.03981  & --0.03078  & --0.11955 & 0.07138 & 0.00549 \\
BWC V33  & 0.55032 & 15.83123 & 0.15391  & --0.06875 & 0.03131 & 0.04492  & --0.01033 & 0.02300 \\
BWC V35  & 0.33048 & 15.61170 & 0.11396  & --0.00048 & 0.01274  & --0.04832 & 0.02117 & 0.00839 \\
BWC V37  & 0.38016 & 15.71013  & --0.08233 & 0.00132 & 0.01138  & --0.10230  & --0.00616 & 0.00039 \\
BWC V41  & 0.46214 & 15.99924 & 0.20650  & --0.09692 & 0.03934 & 0.09486 & 0.01275  & --0.00384 \\
BWC V47  & 0.25692 & 15.76428  & --0.00628 & 0.02310 & 0.00563 & 0.06542 & 0.00325  & --0.00214 \\
BWC V48  & 0.33546 & 15.80626 & 0.11752 & 0.00956 & 0.01069 & 0.07251  & --0.02721  & --0.03532 \\
BWC V51  & 0.64949 & 15.80904 & 0.02438 & 0.00649  & --0.00662 & 0.06748  & --0.01255  & --0.00099 \\
BWC V54  & 0.28870 & 15.88149  & --0.13840  & --0.01482  & --0.00439  & --0.04686  & --0.00739  & --0.01112 \\
BWC V56  & 0.68046 & 15.79886 & 0.04202 & 0.02515 & 0.00362 & 0.07483  & --0.01293  & --0.01304 \\
BWC V59  & 0.26995 & 15.92907 & 0.10475 & 0.00316  & --0.02250 & 0.11769  & --0.02159 & 0.01841 \\
BWC V60  & 0.32069 & 15.90652 & 0.03105 & 0.00408 & 0.01013  & --0.11555 & 0.00931 & 0.00067 \\
BWC V61  & 0.61595 & 15.91442  & --0.08228  & --0.01083 & 0.00406  & --0.04895  & --0.03462  & --0.01377 \\
BWC V62  & 0.28682 & 15.94575 & 0.03045 & 0.02206  & --0.01766  & --0.13968 & 0.00755 & 0.00835 \\
BWC V65  & 0.55720 & 16.25534 & 0.05863 & 0.10905  & --0.07666  & --0.19081 & 0.05050 & 0.04403 \\
BWC V81  & 0.38590 & 16.32912  & --0.02116 & 0.00381 & 0.00326 & 0.04954 & 0.01177 & 0.00647 \\
BWC V106 & 0.46496 & 16.98680 & 0.06253  & --0.01122 & 0.00392 & 0.16846  & --0.03590 & 0.01850 \\
\hline
BW1 V7   & 0.46006 & 15.01576  & --0.02240 & 0.00906 & 0.02357  & --0.05348  & --0.03702  & --0.01078 \\
BW1 V10  & 0.55564 & 15.39061 & 0.12845 & 0.04790  & --0.08271 & 0.17708  & --0.14840 & 0.03186 \\
BW1 V11  & 0.38434 & 15.22302 & 0.04042  & --0.00299  & --0.01707  & --0.12515 & 0.00368  & --0.00575 \\
BW1 V14  & 0.49322 & 15.55832 & 0.03253 & 0.08892 & 0.00747  & --0.21482  & --0.03624 & 0.07540 \\
BW1 V18  & 0.52956 & 15.73140  & --0.13519 & 0.10091 & 0.06874 & 0.20490 & 0.07792  & --0.04383 \\
BW1 V19  & 0.44444 & 15.81823  & --0.25610  & --0.12643  & --0.06836  & --0.09761  & --0.05660  & --0.06691 \\
BW1 V21  & 0.45411 & 15.81379  & --0.01311 & 0.08927  & --0.08576 & 0.24706  & --0.05365  & --0.04995 \\
BW1 V25  & 0.60023 & 15.68415  & --0.06749 & 0.08334 & 0.06161 & 0.16993 & 0.06321  & --0.02440 \\
BW1 V31  & 0.42434 & 16.04257  & --0.10248 & 0.05804 & 0.02722 & 0.18639 & 0.03659  & --0.02078 \\
BW1 V32  & 0.30529 & 15.72653 & 0.03150  & --0.01411  & --0.01006  & --0.13423  & --0.00190  & --0.00208 \\
BW1 V34  & 0.63271 & 15.88880  & --0.08460 & 0.00672 & 0.03775  & --0.13748  & --0.05135  & --0.02145 \\
BW1 V36  & 0.44856 & 16.19532  & --0.28328  & --0.12088  & --0.06208  & --0.02125 & 0.05809 & 0.06630 \\
BW1 V40  & 0.61175 & 15.84933 & 0.01242 & 0.04482 & 0.00367 & 0.12063  & --0.01256  & --0.02590 \\
BW1 V43  & 0.42178 & 16.19170  & --0.15412  & --0.04956  & --0.01251  & --0.15473  & --0.11802  & --0.03584 \\
BW1 V50  & 0.31992 & 16.06068 & 0.09004 & 0.01036  & --0.00873 & 0.07912  & --0.02550 & 0.01269 \\
BW1 V53  & 0.31963 & 16.10365  & --0.10526  & --0.01148  & --0.00306 & 0.07111  & --0.00109 & 0.01793 \\
\hline
\hline
\end{tabular}
}
\end{center}
\pagebreak
 
\begin{center}
Table 3 \\
 
Fourier coefficients for BW2, BW3 and BW4 RR Lyr stars. \\
\vspace{7pt}
 
{\scriptsize
\begin{tabular}{|l|c|c|c|c|c|c|c|c|}
\hline
\hline
 Star & Period & $a_0$ & $a_1$ & $a_2$ & $a_3$ & $b_1$ & $b_2$ & $b_3$
\\
\hline
\hline
BW2 V8   & 0.39402 & 15.30095 & 0.03561 & 0.01842  & --0.04255  & --0.15957 & 0.01801 & 0.00109 \\
BW2 V10  & 0.50784 & 15.370 & --0.00073 & 0.01476 & 0.01702  & --0.11369  & --0.00037  & --0.00203 \\
BW2 V14  & 0.77102 & 15.57770  & --0.09443 & 0.03013 & 0.00661 & 0.15949 & 0.07471 & 0.00726 \\
BW2 V17  & 0.47838 & 15.98977  & --0.24328  & --0.00783 & 0.05207 & 0.08448 & 0.17404 & 0.02971 \\
BW2 V18  & 0.62432 & 15.63196 & 0.04788  & --0.00905 & 0.00174 & 0.01126  & --0.00004  & --0.00367 \\
BW2 V23  & 0.49276 & 16.23642  & --0.00164 & 0.05716 & 0.06564  & --0.23865  & --0.05574 & 0.01775 \\
BW2 V24  & 0.59743 & 15.85332 & 0.07513 & 0.01934  & --0.02928 & 0.09557  & --0.04980 & 0.00552 \\
BW2 V42  & 0.28954 & 16.46074  & --0.06715 & 0.03874 & 0.02568  & --0.14691  & --0.02812 & 0.00881 \\
\hline
BW3 V11  & 0.65494 & 15.35452 & 0.10792 & 0.06812  & --0.04491  & --0.17746 & 0.09222 & 0.02124 \\
BW3 V13  & 0.45751 & 15.71935  & --0.10555 & 0.11798 & 0.04780 & 0.21182 & 0.04801  & --0.05458 \\
BW3 V16  & 0.60354 & 15.66343  & --0.18356  & --0.09563  & --0.05269  & --0.04425  & --0.03095  & --0.02094 \\
BW3 V17  & 0.40333 & 16.03220  & --0.26998  & --0.12746  & --0.08647 & 0.00347 & 0.04934 & 0.08525 \\
BW3 V21  & 0.77304 & 15.68117 & 0.00470 & 0.04666 & 0.00763 & 0.10610 & 0.00311  & --0.01799 \\
BW3 V26  & 0.33435 & 15.831 & --0.06836 & 0.02112  & --0.00734  & --0.11289  & --0.01370 & 0.01559 \\
BW3 V41  & 0.26263 & 16.04087 & 0.01567 & 0.00925  & --0.00481 & 0.13983  & --0.00422  & --0.00801 \\
BW3 V43  & 0.64089 & 15.97926  & --0.03492 & 0.01260  & --0.00494  & --0.07085  & --0.01535 & 0.00590 \\
BW3 V46  & 0.48653 & 16.22392  & --0.07655 & 0.10706 & 0.09037  & --0.19787  & --0.09806 & 0.02662 \\
BW3 V48  & 0.25073 & 16.09169  & --0.03280 & 0.02393 & 0.00320  & --0.08555  & --0.01446 & 0.00233 \\
BW3 V61  & 0.54160 & 16.59679 & 0.12457 & 0.04840  & --0.06556  & --0.16219 & 0.09705  & --0.00834 \\
BW3 V66  & 0.29103 & 16.42890  & --0.04403 & 0.00950  & --0.01221 & 0.12427 & 0.01406 & 0.00046 \\
BW3 V81  & 0.50818 & 16.72673 & 0.11199  & --0.01102  & --0.02497 & 0.08767  & --0.06718 & 0.03610 \\
BW3 V99  & 0.56046 & 17.10945  & --0.00514 & 0.14239 & 0.00074 & 0.22057  & --0.01177  & --0.08083 \\
\hline
BW4 V4   & 0.31997 & 15.03589  & --0.04851  & --0.00473 & 0.00089 & 0.00327  & --0.00121 & 0.00426 \\
BW4 V5   & 0.47468 & 15.43992 & 0.20599  & --0.08595 & 0.04447 & 0.03323 & 0.05749  & --0.07734 \\
BW4 V8   & 0.51586 & 15.47024 & 0.14576  & --0.04536  & --0.01928 & 0.15356  & --0.10313 & 0.07256 \\
BW4 V9   & 0.46317 & 15.60021 & 0.10146 & 0.12321  & --0.00290  & --0.23458 & 0.02154 & 0.09952 \\
BW4 V11  & 0.60169 & 15.33607  & --0.13875  & --0.06246  & --0.03660  & --0.01995 & 0.00198 & 0.01320 \\
BW4 V12  & 0.63958 & 15.44754 & 0.18674  & --0.09813 & 0.06261  & --0.02285 & 0.00211 & 0.00683 \\
BW4 V22  & 0.56093 & 15.80406  & --0.01453 & 0.10079 & 0.06128  & --0.17942  & --0.04395 & 0.05075 \\
BW4 V25  & 0.32191 & 15.73668  & --0.11834 & 0.00052 & 0.00792  & --0.05393  & --0.01552  & --0.00154 \\
BW4 V27  & 0.34744 & 15.70794 & 0.03251  & --0.00295 & 0.00064  & --0.01658 & 0.00323  & --0.00238 \\
BW4 V31  & 0.32701 & 15.84757 & 0.02624 & 0.00802  & --0.01117  & --0.13263 & 0.00574  & --0.00422 \\
BW4 V43  & 0.55992 & 16.07776 & 0.08331 & 0.04332  & --0.03439  & --0.12114 & 0.05432 & 0.01278 \\
\hline
\hline
\end{tabular}
}
\end{center}
\pagebreak
 
\begin{center}
Table 4 \\
 
Fourier coefficients for BW5, BW6, BW7 and BW8 RR Lyr stars. \\
\vspace{7pt}
 
{\scriptsize
\begin{tabular}{|l|c|c|c|c|c|c|c|c|}
\hline
\hline
 Star & Period & $a_0$ & $a_1$ & $a_2$ & $a_3$ & $b_1$ & $b_2$ & $b_3$
\\
\hline
\hline
BW5 V13  & 0.49492 & 15.74369  & --0.22417  & --0.06462  & --0.03345  & --0.02542 & 0.01096 & 0.03057 \\
BW5 V17  & 0.27929 & 15.47744 & 0.07933 & 0.00733  & --0.01147  & --0.11756 & 0.02435 & 0.00171 \\
BW5 V24  & 0.55213 & 15.73153  & --0.14598  & --0.09262  & --0.06636 & 0.09943 & 0.02786  & --0.02964 \\
BW5 V28  & 0.46723 & 16.03722  & --0.08302 & 0.14263  & --0.03555 & 0.24815 & 0.00437  & --0.12243 \\
BW5 V29  & 0.47361 & 16.03377 & 0.23912  & --0.12540 & 0.10197 & 0.07024 & 0.00998  & --0.03506 \\
BW5 V34  & 0.49027 & 16.14363 & 0.23680  & --0.08742 & 0.02517  & --0.02353 & 0.08573  & --0.08483 \\
BW5 V36  & 0.59451 & 15.88070 & 0.08675 & 0.00215  & --0.02257 & 0.10078  & --0.04973 & 0.00802 \\
BW5 V39  & 0.50786 & 16.31342  & --0.21866  & --0.02027 & 0.03603 & 0.08273 & 0.12764 & 0.02889 \\
BW5 V40  & 0.32586 & 16.00048 & 0.12528  & --0.00753 & 0.01011  & --0.03214 & 0.01322  & --0.00217 \\
BW5 V43  & 0.45997 & 15.98198  & --0.02487 & 0.00563  & --0.00023 & 0.02545 & 0.01151  & --0.00123 \\
BW5 V50  & 0.49635 & 16.40812 & 0.00279 & 0.03571  & --0.06864 & 0.22033  & --0.05874  & --0.03346 \\
BW5 V135 & 0.58825 & 17.70118 & 0.07639  & --0.02503 & 0.00666  & --0.02952 & 0.04732  & --0.02411 \\
BW5 V174 & 0.56146 & 18.91344  & --0.23086  & --0.17669  & --0.04552  & --0.09183 & 0.07547 & 0.05427 \\
\hline
BW6 V7   & 0.52501 & 15.43896 & 0.05666 & 0.11906 & 0.02399  & --0.25096  & --0.00951 & 0.06544 \\
BW6 V12  & 0.55603 & 15.62931  & --0.26848 & 0.01874 & 0.11584 & 0.15383 & 0.21035 & 0.10218 \\
BW6 V15  & 0.55745 & 15.52499  & --0.10491  & --0.02593 & 0.01222  & --0.08512  & --0.07019  & --0.03755 \\
BW6 V17  & 0.65163 & 15.58864  & --0.03241 & 0.04355 & 0.01908 & 0.14238 & 0.02348  & --0.02593 \\
BW6 V18  & 0.54140 & 15.92709  & --0.11033 & 0.07300 & 0.07244 & 0.19037 & 0.05496  & --0.01885 \\
BW6 V20  & 0.39370 & 15.95989 & 0.18270  & --0.08668 & 0.02316 & 0.17062  & --0.08077 & 0.07214 \\
BW6 V27  & 0.58400 & 15.75801  & --0.14995  & --0.06970  & --0.03793  & --0.02703  & --0.00286 & 0.00889 \\
BW6 V29  & 0.56290 & 15.89626  & --0.08588 & 0.03675 & 0.03889  & --0.14096  & --0.07449 & 0.00162 \\
BW6 V32  & 0.31333 & 15.82354 & 0.09456  & --0.00719  & --0.00737 & 0.09004  & --0.00417  & --0.00163 \\
BW6 V35  & 0.43124 & 16.24859 & 0.22420  & --0.11523 & 0.06165 & 0.17803  & --0.07829 & 0.08456 \\
BW6 V36  & 0.32026 & 15.86359  & --0.05740 & 0.00588 & 0.00134  & --0.10712  & --0.00609 & 0.01073 \\
BW6 V44  & 0.24855 & 16.01323 & 0.03091 & 0.04952  & --0.00224  & --0.11472 & 0.01885 & 0.00991 \\
BW6 V46  & 0.30942 & 16.07095  & --0.01235 & 0.00807  & --0.01622  & --0.10305  & --0.00009 & 0.00557 \\
\hline
BW7 V8   & 0.50711 & 15.24365  & --0.17698  & --0.01089 & 0.04847  & --0.16168  & --0.13848  & --0.07950 \\
BW7 V15  & 0.49708 & 15.61519  & --0.24593  & --0.04756 & 0.05030 & 0.00903 & 0.09362 & 0.15396 \\
BW7 V18  & 0.66695 & 15.68157  & --0.13058  & --0.08522 & 0.04771  & --0.12733  & --0.06307  & --0.01598 \\
BW7 V20  & 0.77051 & 15.57981  & --0.13360  & --0.01937 & 0.03557  & --0.11187  & --0.10660  & --0.07047 \\
BW7 V23  & 0.60579 & 15.47907 & 0.00493 & 0.01661 & 0.01408 & 0.06425 & 0.00496 & 0.01379 \\
BW7 V24  & 0.26902 & 15.54379  & --0.08860  & --0.02298 & 0.00475  & --0.00269  & --0.00793  & --0.00634 \\
BW7 V25  & 0.52116 & 15.76348  & --0.15565 & 0.02516 & 0.05714 & 0.11551 & 0.09462 & 0.04389 \\
BW7 V30  & 0.36202 & 15.77208  & --0.09676  & --0.00761  & --0.00147 & 0.01737  & --0.00622  & --0.00510 \\
BW7 V31  & 0.29191 & 15.79245 & 0.04443 & 0.01537  & --0.00972 & 0.09211  & --0.01626  & --0.01423 \\
BW7 V33  & 0.51122 & 16.15760 & 0.14645 & 0.067 & --0.07103  & --0.13305 & 0.10726 & 0.03626 \\
BW7 V48  & 0.63862 & 16.36857  & --0.15123  & --0.04067  & --0.04215 & 0.02684 & 0.08024 & 0.05194 \\
BW7 V51  & 0.27215 & 16.23314 & 0.13253 & 0.00308  & --0.01010 & 0.03316  & --0.03339 & 0.00837 \\
\hline
BW8 V7   & 0.55454 & 15.19504  & --0.18544  & --0.09369  & --0.04351  & --0.05748  & --0.03368  & --0.06820 \\
BW8 V8   & 0.78117 & 15.13834  & --0.06450 & 0.04008  & --0.00211  & --0.15251  & --0.03431 & 0.03097 \\
BW8 V15  & 0.57724 & 15.43612 & 0.10483  & --0.03919  & --0.00381 & 0.12081  & --0.06165 & 0.02971 \\
BW8 V16  & 0.42413 & 15.63486 & 0.24479  & --0.14020 & 0.06679 & 0.04301 & 0.05651  & --0.04042 \\
BW8 V18  & 0.70001 & 15.36377 & 0.02815 & 0.03650  & --0.03597 & 0.13366  & --0.01826  & --0.01448 \\
BW8 V20  & 0.67699 & 15.37804 & 0.07722 & 0.01292  & --0.01311  & --0.10345 & 0.05424  & --0.01365 \\
BW8 V26  & 0.61699 & 15.48143 & 0.01759 & 0.01907  & --0.00690  & --0.06553  & --0.00096  & --0.00086 \\
BW8 V28  & 0.53525 & 15.88063 & 0.22784  & --0.08645 & 0.04224  & --0.05574 & 0.08388  & --0.06101 \\
BW8 V34  & 0.32037 & 15.69600 & 0.04183  & --0.01244  & --0.00256  & --0.11836  & --0.00027  & --0.00187 \\
BW8 V35  & 0.64770 & 15.63083  & --0.03268 & 0.00110 & 0.00622  & --0.04185  & --0.01501 & 0.00548 \\
BW8 V36  & 0.28612 & 15.78090 & 0.00710  & --0.00267  & --0.01242  & --0.14495 & 0.00500 & 0.00697 \\
\hline
\hline
\end{tabular}
}
\end{center}
\pagebreak
 
\begin{center}
Table 5 \\
 
Fourier coefficients for BW9, BW10 and BW11 RR Lyr stars. \\
\vspace{7pt}
 
{\scriptsize
\begin{tabular}{|l|c|c|c|c|c|c|c|c|}
\hline
\hline
 Star & Period & $a_0$ & $a_1$ & $a_2$ & $a_3$ & $b_1$ & $b_2$ & $b_3$
\\
\hline
\hline
BW9 V14  & 0.60755 & 15.67594  & 0.03750  & 0.18716  & 0.04929  &  --0.265 &  --0.00639  & 0.12424\\
BW9 V15  & 0.48592 & 15.81885  & 0.05686  & 0.08996  &  --0.11034  & 0.25305  &  --0.11916  &  --0.00948\\
BW9 V22  & 0.63233 & 15.70985  &  --0.01539  & 0.06115  & 0.02738  & 0.155 & 0.00343  &  --0.03450\\
BW9 V24  & 0.47635 & 15.89431  &  --0.05614  & 0.05746  & 0.01684  &  --0.24337  &  --0.109 &  --0.00866\\
BW9 V34  & 0.44930 & 16.09160  &  --0.00948  & 0.11077  &  --0.05787  & 0.26449  &  --0.05935  &  --0.01167\\
BW9 V35  & 0.59252 & 15.94749  &  --0.07101  & 0.06420  & 0.02670  &  --0.14381  &  --0.05221  & 0.01081\\
BW9 V37  & 0.54705 & 16.03117  & 0.14788  &  --0.06232  &  --0.02678  & 0.09852  &  --0.10603  & 0.07518\\
BW9 V38  & 0.30574 & 15.88839  &  --0.01008  & 0.01025  &  --0.02362  &  --0.11762  & 0.00685  & 0.00304\\
BW9 V43  & 0.33260 & 15.98245  &  --0.13432  &  --0.01537  &  --0.00434  & 0.02014  & 0.02522  & 0.03012\\
BW9 V51  & 0.67741 & 16.22408  &  --0.04713  & 0.05804  &  --0.02025  & 0.16145  & 0.01879  &  --0.04175\\
BW9 V52  & 0.34218 & 16.07193  &  --0.10525  &  --0.01760  & 0.04511  & 0.10789  &  --0.01454  & 0.04604\\
BW9 V54  & 0.54766 & 16.25378  & 0.08813  & 0.13094  &  --0.10769  &  --0.21414  & 0.07819  & 0.03125\\
BW9 V55  & 0.42564 & 16.37788  & 0.08224  & 0.07466  &  --0.25121  & 0.35862  &  --0.12076  & 0.07347\\
BW9 V59  & 0.54499 & 16.42001  & 0.07427  & 0.14450  & 0.00596  & 0.24228  &  --0.08669  &  --0.08117\\
BW9 V114 & 0.20767 & 16.90682  &  --0.10789  &  --0.01132  &  --0.00158  &  --0.02055  &  --0.01153  &  --0.01480\\
BW9 V215 & 0.30284 & 17.82195  &  --0.06458  & 0.00055  &  --0.00796  &  --0.04049  &  --0.00191  & 0.00621\\
\hline
BW10 V8  & 0.71337 & 15.10997  &  --0.13883  & 0.04548  & 0.06250  & 0.19182  & 0.10997  &  --0.00180\\
BW10 V14 & 0.48410 & 15.43848  & 0.22086  &  --0.12785  & 0.05796  & 0.08062  &  --0.05926  & 0.06985\\
BW10 V20 & 0.52711 & 15.74564  &  --0.15945  & 0.08759  & 0.03315  & 0.19493  & 0.08116  &  --0.06317\\
BW10 V21 & 0.52204 & 15.49528  &  --0.12522  &  --0.04779  &  --0.01772  &  --0.00815  &  --0.02538  &  --0.00654\\
BW10 V27 & 0.25828 & 15.60068  & 0.03626  & 0.04846  &  --0.02555  & 0.13236  &  --0.02561  &  --0.00476\\
BW10 V36 & 0.27700 & 15.66528  &  --0.08551  &  --0.00145  &  --0.01159  & 0.02356  & 0.00473  &  --0.02121\\
BW10 V39 & 0.51363 & 16.01677  & 0.12177  & 0.05907  &  --0.04160  &  --0.13580  & 0.04981  & 0.01644\\
BW10 V40 & 0.68277 & 15.75784  & 0.04401  & 0.06504  &  --0.01188  & 0.12006  &  --0.02501  &  --0.01914\\
BW10 V41 & 0.62690 & 15.97526  & 0.19554  & 0.00727  & 0.00064  &  --0.09987  & 0.11692  & 0.00322\\
BW10 V44 & 0.52727 & 15.97825  &  --0.16906  &  --0.05581  &  --0.03842  &  --0.06636  &  --0.06060  &  --0.01437\\
BW10 V45 & 0.25283 & 15.82750  & 0.08871  &  --0.02732  & 0.01504  & 0.07542  &  --0.00145  &  --0.00738\\
BW10 V46 & 0.32542 & 15.91897  &  --0.14226  &  --0.00847  & 0.01862  &  --0.04850  &  --0.04286  &  --0.03933\\
BW10 V48 & 0.32333 & 15.90164  &  --0.13477  &  --0.00085  & 0.02157  &  --0.03972  &  --0.06821  &  --0.05217\\
BW10 V51 & 0.27651 & 15.96364  &  --0.08011  &  --0.00689  & 0.00413  &  --0.07544  &  --0.00740  &  --0.00159\\
BW10 V56 & 0.61826 & 15.97134  &  --0.07741  &  --0.02066  &  --0.00403  & 0.00611  &  --0.00276  & 0.00799\\
BW10 V59 & 0.31930 & 16.05074  &  --0.06682  & 0.00278  & 0.01417  & 0.10301  & 0.01773  & 0.01681\\
BW10 V60 & 0.30371 & 16.15556  & 0.14915  &  --0.01673  & 0.01003  & 0.01147  & 0.01265  & 0.00661\\
BW10 V65 & 0.28099 & 16.15793  &  --0.03777  & 0.04037  & 0.04111  & 0.16940  & 0.01602  &  --0.01399\\
BW10 V66 & 0.58123 & 16.35251  &  --0.10078  & 0.02382  & 0.03801  & 0.12306  & 0.05516  &  --0.00723\\
BW10 V70 & 0.28301 & 16.21685  & 0.14472  &  --0.02278  & 0.00709  & 0.02446  & 0.00599  & 0.01668\\
BW10 V95 & 0.63810 & 16.66445  & 0.06559  & 0.02317  &  --0.03673  & 0.11754  &  --0.05475  &  --0.00215\\
\hline
BW11 V3  & 0.59598 & 14.57884  &  --0.00511  & 0.03404  &  --0.03224  &  --0.21365  &  --0.05201  & 0.02121\\
BW11 V10 & 0.45066 & 15.65054  & 0.08474  & 0.04920  & 0.02891  &  --0.23224  &  --0.02664  & 0.06349\\
BW11 V17 & 0.49491 & 15.85333  & 0.12509  & 0.12631  & 0.01748  &  --0.25723  & 0.04715  & 0.11588\\
BW11 V23 & 0.27402 & 15.62907  &  --0.12778  &  --0.01440  &  --0.02054  & 0.05640  & 0.01454  &  --0.03761\\
BW11 V25 & 0.51660 & 15.96370  & 0.32168  &  --0.16352  & 0.03204  & 0.02419  & 0.16834  &  --0.11914\\
BW11 V29 & 0.46610 & 16.05572  & 0.19887  & 0.00055  &  --0.05268  &  --0.12174  & 0.12812  &  --0.02661\\
BW11 V34 & 0.34167 & 15.83525  &  --0.04116  & 0.02383  & 0.01017  & 0.12315  & 0.01428  &  --0.00247\\
BW11 V36 & 0.51394 & 16.11506  & 0.13526  &  --0.05910  & 0.01098  & 0.14378  &  --0.06355  & 0.05943\\
BW11 V38 & 0.59301 & 15.91869  & 0.05810  & 0.04109  &  --0.03999  &  --0.12999  & 0.04167  & 0.00375\\
BW11 V39 & 0.31226 & 15.90271  & 0.03575  & 0.03137  &  --0.01076  &  --0.13420  & 0.01516  & 0.00830\\
BW11 V44 & 0.48248 & 16.26245  &  --0.12438  & 0.04703  & 0.07152  & 0.16913  & 0.06336  & 0.00376\\
BW11 V55 & 0.28618 & 16.24522  &  --0.00384  &  --0.01892  & 0.02958  &  --0.12525  & 0.00902  & 0.00422\\
\hline
\hline
\end{tabular}
}
\end{center}
\pagebreak

\begin{center}
Table 6 \\
 
Fourier coefficients for MM5A, MM5B, MM7A and MM7B RR Lyr stars. \\
\vspace{7pt}
 
{\scriptsize
\begin{tabular}{|l|c|c|c|c|c|c|c|c|}
\hline
\hline
 Star & Period & $a_0$ & $a_1$ & $a_2$ & $a_3$ & $b_1$ & $b_2$ & $b_3$
\\
\hline
\hline
MM5-A V6 & 0.51202 & 15.49398  & --0.21193  & --0.06483  & --0.03117 & 0.05239 & 0.10877 & 0.01178\\
MM5-A V9 & 0.51619 & 15.70435 & 0.26126  & --0.09831 & 0.03567  & --0.00672 & 0.07100  & --0.09076\\
MM5-A V15& 0.60022 & 15.72056 & 0.09248 & 0.05317  & --0.05042  & --0.14779 & 0.07951 & 0.00156\\
MM5-A V18& 0.52667 & 15.96786  & --0.16892 & 0.01696 & 0.06189 & 0.14879 & 0.09686 & 0.03134\\
MM5-A V19& 0.61251 & 15.92820 & 0.08290 & 0.09745  & --0.06584 & 0.20467  & --0.09701  & --0.03419\\  
MM5-A V20& 0.39119 & 15.70409  & --0.10723  & --0.00904 & 0.01498  & --0.07150 & 0.00763  & --0.01238\\
MM5-A V21& 0.46382 & 16.11532  & --0.09888 & 0.05533  & --0.00188 & 0.25375  & --0.00806  & --0.08262\\
MM5-A V27& 0.62157 & 16.00873 & 0.09686  & --0.02810 & 0.01729  & --0.02969 & 0.02812 & 0.01586\\
MM5-A V32& 0.54469 & 16.31287 & 0.11932  & --0.00098 & 0.00096  & --0.10804 & 0.04829 & 0.00440\\
MM5-A V37& 0.45791 & 16.61944 & 0.08044 & 0.08155  & --0.12812 & 0.26779  & --0.12842  & --0.00102\\
MM5-A V41& 0.58450 & 16.55360  & --0.22431  & --0.16234  & --0.10492  & --0.10453  & --0.00709 & 0.06610\\
MM5-A V46& 0.37019 & 16.52867 & 0.19783  & --0.06472 & 0.02040 & 0.02212  & --0.02326 & 0.00951\\
\hline
MM5-B V4 & 0.52489 & 15.20260 & 0.16128  & --0.06602 & 0.00139 & 0.10494  & --0.06819 & 0.07335\\
MM5-B V10& 0.52391 & 15.80769 & 0.21502 & 0.01624  & --0.06884  & --0.12193 & 0.13659  & --0.05945\\
MM5-B V20& 0.64994 & 15.87767 & 0.10679  & --0.03176  & --0.01329  & --0.11265 & 0.05518  & --0.03786\\
MM5-B V21& 0.50359 & 16.10172 & 0.00034 & 0.13380 & 0.04152 & 0.23838  & --0.04565  & --0.08602\\
MM5-B V28& 0.47995 & 16.26156 & 0.04026 & 0.04832  & --0.09955 & 0.24168  & --0.08479 & 0.00556\\
MM5-B V31& 0.36456 & 16.47650 & 0.12577 & 0.00712  & --0.06433 & 0.23951  & --0.16559 & 0.06840\\
MM5-B V41& 0.25745 & 16.29955 & 0.08252 & 0.01434  & --0.02669  & --0.11422 & 0.02376  & --0.01327\\
MM5-B V44& 0.52311 & 16.66455 & 0.20763  & --0.08950 & 0.04017 & 0.13707  & --0.08540 & 0.07512\\
\hline
MM7-A V2 & 0.54078 & 15.13520 & 0.17003  & --0.06055  & --0.00151 & 0.15396  & --0.08457 & 0.06661\\
MM7-A V16& 0.56710 & 15.87458  & --0.09581 & 0.09339 & 0.08503 & 0.20386 & 0.04160  & --0.02872\\
MM7-A V20& 0.52984 & 16.02787  & --0.24720  & --0.14335  & --0.07130  & --0.13666  & --0.06001  & -0.08389\\
MM7-A V25& 0.49380 & 15.96739 & 0.07652 & 0.00610  & --0.05939 & 0.21915  & --0.10446 & 0.01255\\
MM7-A V26& 0.44361 & 16.16516 & 0.25202  & --0.08683 & 0.05675  & --0.00917 & 0.02388  & --0.03266\\
MM7-A V34& 0.53049 & 16.05563  & --0.01475 & 0.03073 & 0.00438  & --0.17186  & --0.04794 & 0.02782\\
MM7-A V44& 0.22759 & 16.12636 & 0.12231 & --0.01129 & --0.00137 & --0.03649 & 0.01418 & --0.03026\\ 
MM7-A V47& 0.28782 & 16.20053 & 0.11638 & 0.01052 & 0.00575  & --0.08639 & 0.03521 & 0.00012\\
MM7-A V94& 0.27499 & 17.03946 & 0.01557 & 0.01524 & 0.01033 & 0.11992 & 0.00193  & --0.01247\\
\hline
MM7-B V6 & 0.53913 & 15.66624 & 0.01717 & 0.06364 & 0.01106  & --0.19809  & --0.03479 & 0.05570\\
MM7-B V7 & 0.59002 & 15.73317 & 0.02054 & 0.08373  & --0.06861  & --0.20587 & 0.00904 & 0.05533\\
MM7-B V8 & 0.55940 & 15.69437  & --0.04981 & 0.09440 & 0.01978 & 0.17077 & 0.02745  & --0.05198\\
MM7-B V12& 0.88121 & 15.69762  & --0.06280  & --0.00535  & --0.00768 & 0.08718 & 0.03442 & 0.00477\\
MM7-B V13& 0.53321 & 15.88185  & --0.14795  & --0.06607  & --0.02875  & --0.11044  & --0.04872  & -0.02883\\
MM7-B V14& 0.28675 & 15.76076 & 0.07046  & --0.00291 & 0.01445  & --0.09978 & 0.01526 & 0.01005\\
MM7-B V63& 0.61806 & 17.39339 & 0.24643  & --0.11844 & 0.04429  & --0.04129 & 0.08967  & --0.04796\\
\hline
\hline
\end{tabular}
}
\end{center}
\pagebreak
\begin{center}
Table 7 \\
 
Parameters of RR Lyr stars from Baade's Window. \\
\vspace{7pt}
 
{\scriptsize 
\begin{tabular}{|l|c|c|c|c|c|c|c|}
\hline
\hline
Star & Period & $\langle I \rangle$ & $\langle V-I \rangle$ & $I_0$ &
$(V-I)_0$ & $V_{V-I}$ & Type \\
\hline
\hline
BWC V6    & 0.42765 & 15.172  & 1.215 & 14.289  &  0.629  &  13.351 & RRab$^*$ \\
BWC V14   & 0.44022 & 15.719  & 0.904 & 14.841  &  0.311  &  14.363 & RRab$^*$ \\
BWC V15   & 0.45871 & 15.918  & 0.807 & 14.928  &  0.151  &  14.708 & RRab$^*$ \\
BWC V17   & 0.29872 & 15.339  & 0.939 &   --   &    --   &   13.929 & RRc \\
BWC V22   & 0.48968 & 15.762  & 1.171 & 14.734  &  0.503  &  14.005 & RRab$^*$ \\
BWC V23   & 0.45426 & 15.807  & 1.028 & 14.941  &  0.444  &  14.265 & RRab$^*$ \\
BWC V25   & 0.47418 & 15.728  & 1.063 &   --   &    --   &   14.133 & RRab \\
BWC V26   & 0.47863 & 16.012  & 1.273 & 15.120  &  0.669  &  14.102 & RRab$^*$ \\
BWC V28   & 0.59478 & 15.625  & 1.262 & 14.623  &  0.604  &  13.732 & RRab$^*$ \\
BWC V30   & 0.57147 & 15.667  & 1.164 & 14.659  &  0.482  &  13.921 & RRab$^*$ \\
BWC V33   & 0.55032 & 15.832  & 1.224 & 14.942  &  0.618  &  13.996 & RRab$^*$ \\
BWC V35   & 0.33048 & 15.611  & 1.063 & 14.600  &  0.398  &  14.017 & RRc \\
BWC V37   & 0.38016 & 15.710  & 1.023 & 14.727  &  0.362  &  14.175 & RRc \\
BWC V41   & 0.46214 & 15.998  & 1.325 & 14.923  &  0.626  &  14.010 & RRab$^*$ \\
BWC V47   & 0.25692 & 15.829  & 0.800 &   --   &    --   &   14.629 & RRc \\
BWC V48   & 0.33546 & 15.809  & 1.124 & 14.870  &  0.480  &  14.123 & RRc \\
BWC V51   & 0.64949 & 15.806  & 1.286 & 14.875  &  0.666  &  13.876 & RRab$^*$ \\
BWC V54   & 0.28870 & 15.916  & 0.958 &   --   &    --   &   14.479 & RRc \\
BWC V56   & 0.68046 & 15.796  & 1.488 & 14.633  &  0.726  &  13.564 & RRab$^*$ \\
BWC V59   & 0.26995 & 15.928  & 0.900 &   --   &    --   &   14.578 & RRc \\
BWC V60   & 0.32069 & 15.907  & 1.047 & 14.972  &  0.425  &  14.336 & RRc \\
BWC V61   & 0.61595 & 15.915  & 1.350 & 14.831  &  0.639  &  13.890 & RRab$^*$ \\
BWC V62   & 0.28682 & 15.946  & 1.135 &   --   &    --   &   14.243 & RRc \\
BWC V65   & 0.55720 & 16.249  & 0.507 & 15.231  & --0.179  &  15.488 & RRab$^*$ \\
BWC V81   & 0.38590 & 16.331  & 1.178 & 15.459  &  0.599  &  14.564 & RRc \\
BWC V106  & 0.46496 & 16.987  & 1.508 & 15.943  &  0.826  &  14.726 & RRab$^*$ \\
\hline
BW1 V7    & 0.46006 & 15.143  & 1.659 & 14.267  &  1.077  &  12.654 & RRab$^*$ \\
BW1 V10   & 0.55564 & 15.397  & 1.045 & 14.468  &  0.429  &  13.830 & RRab$^*$ \\
BW1 V11   & 0.38434 & 15.222  & 1.264 & 14.087  &  0.530  &  13.326 & RRc \\
BW1 V14   & 0.49322 & 15.551  & 1.170 & 14.706  &  0.595  &  13.796 & RRab$^*$ \\
BW1 V18   & 0.52956 & 15.730  & 1.111 & 14.756  &  0.456  &  14.062 & RRab$^*$ \\
BW1 V19   & 0.44444 & 15.816  & 1.339 & 14.826  &  0.683  &  13.807 & RRab$^*$ \\
BW1 V21   & 0.45411 & 15.820  & 1.233 & 14.814  &  0.555  &  13.970 & RRab$^*$ \\
BW1 V25   & 0.60023 & 15.688  & 1.247 & 14.704  &  0.595  &  13.818 & RRab$^*$ \\
BW1 V31   & 0.42434 & 16.026  & 1.189 &   --   &    --   &   14.242 & RRab \\
BW1 V32   & 0.30529 & 15.726  & 0.967 & 14.833  &  0.387  &  14.275 & RRc \\
BW1 V34   & 0.63271 & 15.894  & 1.223 & 14.957  &  0.598  &  14.059 & RRab$^*$ \\
BW1 V36   & 0.44856 & 16.204  & 1.326 & 15.067  &  0.567  &  14.215 & RRab$^*$ \\
BW1 V40   & 0.61175 & 15.847  & 1.256 & 14.859  &  0.636  &  13.964 & RRab$^*$ \\
BW1 V43   & 0.42178 & 16.186  & 1.017 & 15.327  &  0.444  &  14.661 & RRab$^*$ \\
BW1 V50   & 0.31992 & 16.252  & 0.909 &   --   &    --   &   14.889 & RRc \\
BW1 V53   & 0.31963 & 16.104  & 1.138 & 15.072  &  0.442  &  14.397 & RRc \\
\hline
\hline
\end{tabular}
}
\pagebreak
 
Table 7 \\
 
Continued \\
\vspace{7pt}
 
{\scriptsize
 
\begin{tabular}{|l|c|c|c|c|c|c|c|}
\hline
\hline
Star & Period & $\langle I \rangle$ & $\langle V-I \rangle$ & $I_0$ &
$(V-I)_0$ & $V_{V-I}$ & Type \\
\hline
\hline
BW2 V8    & 0.39402 & 15.300  & 1.171 &  14.257 &  0.472 & 13.544 & RRc \\
BW2 V10   & 0.50784 & 15.370  & 1.050 &  14.295 &  0.360 & 13.796 & RRc \\
BW2 V14   & 0.77102 & 15.578  &   --   &    --  &    --  &   --   & RRab \\
BW2 V17   & 0.47838 & 15.977  &   --   &    --  &    --  &   --   & RRab \\
BW2 V18   & 0.62432 & 15.770  & 1.139 &  14.764 &  0.501 & 14.061 & RRc \\
BW2 V23   & 0.49276 & 16.241  & 0.990 &  15.188 &  0.294 & 14.757 & RRab$^*$ \\
BW2 V24   & 0.59743 & 15.852  & 1.436 &  14.689 &  0.684 & 13.697 & RRab$^*$ \\
BW2 V42   & 0.28954 & 16.462  & 1.038 &  15.424 &  0.366 & 14.904 & RRc \\
\hline
BW3 V11   & 0.65494 & 15.352  & 1.372 &  14.030 &  0.491 & 13.294 & RRab$^*$ \\
BW3 V13   & 0.45751 & 15.719  & 1.131 &  14.719 &  0.488 & 14.022 & RRab$^*$ \\
BW3 V16   & 0.60354 & 15.745  & 1.289 &  14.477 &  0.445 & 13.811 & RRab$^*$ \\
BW3 V17   & 0.40333 & 16.039  & 1.167 &  14.982 &  0.470 & 14.288 & RRab$^*$ \\
BW3 V21   & 0.77304 & 15.681  & 1.292 &  14.599 &  0.589 & 13.743 & RRab$^*$ \\
BW3 V26   & 0.33435 & 15.830  & 1.040 &  14.791 &  0.371 & 14.270 & RRc \\
BW3 V41   & 0.26263 & 16.042  & 1.120 &  14.884 &  0.366 & 14.362 & RRc \\
BW3 V43   & 0.64089 & 15.981  & 1.514 &  14.616 &  0.621 & 13.709 & RRab$^*$ \\
BW3 V46   & 0.48653 & 16.226  & 1.335 &  14.995 &  0.499 & 14.223 & RRab$^*$ \\
BW3 V48   & 0.25073 & 16.094  & 1.159 &  14.962 &  0.412 & 14.355 & RRc \\
BW3 V61   & 0.54160 & 16.594  & 1.216 &  15.604 &  0.553 & 14.770 & RRab$^*$ \\
BW3 V66   & 0.29103 & 16.428  & 1.171 &  15.248 &  0.386 & 14.671 & RRc \\
BW3 V81   & 0.50818 & 16.731  &   --   &    --  &    --  &   --   & RRab \\
BW3 V99   & 0.56046 & 17.110  & 1.488 &  15.435 &  0.384 & 14.877 & RRab$^*$ \\
\hline
BW4 V4    & 0.31997 & 15.035  & 1.110 &  13.864 &  0.330 & 13.370 & RRc \\
BW4 V5    & 0.47468 & 15.437  & 1.310 &  14.429 &  0.607 & 13.472 & RRab$^*$ \\
BW4 V8    & 0.51586 & 15.464  & 1.155 &  14.536 &  0.527 & 13.731 & RRab$^*$ \\
BW4 V9    & 0.46317 & 15.602  & 1.220 &  14.612 &  0.540 & 13.773 & RRab$^*$ \\
BW4 V11   & 0.60169 & 15.338  & 1.159 &  14.400 &  0.517 & 13.600 & RRab$^*$ \\
BW4 V12   & 0.63958 & 15.489  & 1.301 &  14.448 &  0.598 & 13.537 & RRab$^*$ \\
BW4 V22   & 0.56093 & 15.799  & 1.408 &  14.780 &  0.687 & 13.687 & RRab$^*$ \\
BW4 V25   & 0.32191 & 15.737  & 1.087 &  14.668 &  0.382 & 14.106 & RRc \\
BW4 V27   & 0.34744 & 15.706  & 1.425 &  14.683 &  0.745 & 13.568 & RRc \\
BW4 V31   & 0.32701 & 15.848  & 1.075 &    --  &    --  &  14.236 & RRc \\
BW4 V43   & 0.55992 & 16.080  & 1.267 &  15.079 &  0.586 & 14.180 & RRab$^*$ \\
BW4 V46   & 0.26256 & 16.067  & 1.022 &  15.146 &  0.387 & 14.533 & RRc \\
BW4 V57   & 0.29085 & 16.490  & 0.769 &  15.526 &  0.111 & 15.337 & RRc \\
BW4 V132  & 0.65076 & 17.412  & 1.248 &  16.504 &  0.623 & 15.540 & RRab \\
\hline
\hline
\end{tabular}
}
\pagebreak
 
Table 7  \\
 
Completed \\
\vspace{7pt}
 
{\scriptsize
 
\begin{tabular}{|l|c|c|c|c|c|c|c|}
\hline
\hline
Star & Period & $\langle I \rangle$ & $\langle V-I \rangle$ & $I_0$ &
$(V-I)_0$ & $V_{V-I}$ & Type \\
\hline
\hline
BW5 V13   & 0.49492 & 15.745  & 0.997 & 14.856  & 0.390  &14.250 & RRab$^*$ \\
BW5 V17   & 0.27929 & 15.478  & 1.000 & 14.446  & 0.321  &13.978 & RRc \\
BW5 V24   & 0.55213 & 15.733  & 1.092 & 14.795  & 0.469  &14.096 & RRab$^*$ \\
BW5 V28   & 0.46723 & 16.032  & 1.086 & 15.061  & 0.453  &14.403 & RRab$^*$ \\
BW5 V29   & 0.47361 & 16.029  & 1.153 & 14.874  & 0.384  &14.299 & RRab$^*$ \\
BW5 V34   & 0.49027 & 16.206  & 1.267 & 15.078  & 0.506  &14.306 & RRab$^*$ \\
BW5 V36   & 0.59451 & 15.928  & 1.414 & 14.842  & 0.663  &13.806 & RRab$^*$ \\
BW5 V39   & 0.50786 & 16.313  & 1.471 & 15.177  & 0.709  &14.106 & RRab$^*$ \\
BW5 V40   & 0.32586 & 16.029  & 1.286 & 14.933  & 0.561  &14.100 & RRc \\
BW5 V43   & 0.45997 & 15.981  & 1.425 & 14.912  & 0.708  &13.844 & RRc \\
BW5 V50   & 0.49635 & 16.407  & 1.176 & 15.400  & 0.498  &14.643 & RRab$^*$ \\
BW5 V135  & 0.58825 & 17.700  & 1.707 &    --  &     --  &  15.140 & RRab \\
BW5 V174  & 0.56146 & 18.956  & 1.066 & 17.900  & 0.365  &17.356 & RRab \\
\hline
BW6 V7    & 0.52501 & 15.442  & 1.164 & 14.404  & 0.468  &13.696 & RRab$^*$ \\
BW6 V12   & 0.55603 & 15.622  & 1.295 & 14.584  & 0.613  &13.680 & RRab$^*$ \\
BW6 V15   & 0.55745 & 15.527  & 1.661 & 14.490  & 1.011  &13.036 & RRab$^*$ \\
BW6 V17   & 0.65163 & 15.584  & 1.336 & 14.546  & 0.680  &13.581 & RRab$^*$ \\
BW6 V18   & 0.54140 & 15.933  & 1.401 & 14.777  & 0.639  &13.831 & RRab$^*$ \\
BW6 V20   & 0.39370 & 15.964  & 1.232 & 14.938  & 0.560  &14.117 & RRab$^*$ \\
BW6 V27   & 0.58400 & 15.760  & 1.517 & 14.610  & 0.754  &13.485 & RRab$^*$ \\
BW6 V29   & 0.56290 & 15.899  & 1.533 &    --  &     --  &  13.600 & RRab \\
BW6 V32   & 0.31333 & 15.823  & 1.092 & 14.802  & 0.444  &14.185 & RRc \\
BW6 V35   & 0.43124 & 16.236  & 1.364 & 15.127  & 0.634  &14.190 & RRab$^*$ \\
BW6 V36   & 0.32026 & 15.953  & 1.109 & 14.917  & 0.419  &14.290 & RRc \\
BW6 V44   & 0.24855 & 16.195  & 0.871 & 15.170  & 0.217  &14.890 & RRc \\
BW6 V46   & 0.30942 & 16.073  & 1.144 & 15.004  & 0.440  &14.357 & RRc \\
\hline
BW7 V8    & 0.50711 & 15.293  & 1.056 & 14.301  & 0.402  &13.709 & RRab$^*$ \\
BW7 V15   & 0.49708 & 15.664  & 1.136 & 14.753  & 0.530  &13.960 & RRab$^*$ \\
BW7 V18   & 0.66695 & 15.672  & 1.215 & 14.389  & 0.393  &13.850 & RRab$^*$ \\
BW7 V20   & 0.77051 & 15.590  & 1.155 & 14.617  & 0.511  &13.857 & RRab$^*$ \\
BW7 V23   & 0.60579 & 15.478  & 1.273 & 14.506  & 0.632  &13.567 & RRab$^*$ \\
BW7 V24   & 0.26902 & 15.542  & 1.229 &    --  &     --  &  13.699 & RRc \\
BW7 V25   & 0.52116 & 15.775  & 1.223 & 14.766  & 0.553  &13.941 & RRab$^*$ \\
BW7 V30   & 0.36202 & 15.768  & 1.125 & 14.790  & 0.486  &14.080 & RRc \\
BW7 V31   & 0.29191 & 15.972  & 0.920 & 14.894  & 0.222  &14.592 & RRc \\
BW7 V33   & 0.51122 & 16.160  & 1.761 & 14.746  & 0.797  &13.518 & RRab$^*$ \\
BW7 V48   & 0.63862 & 16.376  & 1.318 & 15.301  & 0.616  &14.399 & RRab$^*$ \\
BW7 V51   & 0.27215 & 16.591  & 1.029 & 15.177  & 0.060  &15.049 & RRc \\
\hline
BW8 V7    & 0.55454 & 15.199  & 1.107 & 14.280  & 0.493  &13.539 & RRab$^*$ \\
BW8 V8    & 0.78117 & 15.137  & 1.187 & 14.179  & 0.524  &13.357 & RRab$^*$ \\
BW8 V15   & 0.57724 & 15.432  & 1.109 & 14.593  & 0.558  &13.768 & RRab$^*$ \\
BW8 V16   & 0.42413 & 15.633  & 1.107 & 14.775  & 0.531  &13.972 & RRab$^*$ \\
BW8 V18   & 0.70001 & 15.367  & 1.248 & 14.385  & 0.570  &13.495 & RRab$^*$ \\
BW8 V20   & 0.67699 & 15.378  & 1.132 & 14.485  & 0.547  &13.679 & RRab$^*$ \\
BW8 V26   & 0.61699 & 15.483  & 1.176 & 14.593  & 0.588  &13.720 & RRab$^*$ \\
BW8 V28   & 0.53525 & 15.876  & 1.469 &    --  &     --  &  13.674 & RRab \\
BW8 V34   & 0.32037 & 15.699  & 0.946 & 14.875  & 0.390  &14.280 & RRc \\
BW8 V35   & 0.64770 & 15.632  & 1.249 & 14.699  & 0.634  &13.758 & RRab$^*$ \\
BW8 V36   & 0.28612 & 15.779  & 0.971 & 14.913  & 0.408  &14.322 & RRc \\
\hline
\hline
\end{tabular}
}
\end{center}
\pagebreak
 
\begin{center}
Table 8 \\
 
Parameters of RR Lyr stars from BW9, BW10 and BW11 fields. \\
\vspace{7pt}
 
{\scriptsize
\begin{tabular}{|l|c|c|c|c|c|}
\hline
Star & Period & $\langle I \rangle$ & $\langle V-I \rangle$ & $V_{V-I}$
 & Type \\
\hline
BW9 V14    & 0.60755  &15.671  &  1.443  &13.507  & RRab$^*$\\
BW9 V15    & 0.48592  &16.007  &  1.182  &14.234  & RRab$^*$\\
BW9 V22    & 0.63233  &15.711  &  1.466  &13.512  & RRab$^*$\\
BW9 V24    & 0.47635  &15.901  &  1.450  &13.727  & RRab$^*$\\
BW9 V34    & 0.44930  &16.089  &  1.404  &13.983  & RRab$^*$\\
BW9 V35    & 0.59252  &15.954  &  1.548  &13.632  & RRab$^*$\\
BW9 V37    & 0.54705  &16.022  &  1.402  &13.920  & RRab$^*$\\
BW9 V38    & 0.30574  &15.891  &  1.292  &13.952  & RRc\\
BW9 V43    & 0.33260  &15.985  &  1.586  &13.606  & RRc\\
BW9 V51    & 0.67741  &16.586  &  1.023  &15.051  & RRab$^*$\\
BW9 V52    & 0.34218  &16.078  &  1.316  &14.104  & RRc\\
BW9 V54    & 0.54766  &16.250  &  1.575  &13.887  & RRab$^*$\\
BW9 V55    & 0.42564  &16.411  &  1.412  &14.293  & RRab$^*$\\
BW9 V59    & 0.54499  &16.417  &  1.460  &14.226  & RRab$^*$\\
BW9 V114   & 0.20767  &16.906  &  1.460  &14.716  & RRc\\
BW9 V215   & 0.30284  &17.821  &  1.401  &15.719  & RRc\\
\hline
BW10 V8    & 0.71337  &15.107  &    --    &  --     & RRab\\
BW10 V14   & 0.48410  &15.436  &  1.784  &12.760  & RRab$^*$\\
BW10 V20   & 0.52711  &15.749  &  1.381  &13.678  & RRab$^*$\\
BW10 V21   & 0.52204  &15.497  &  1.388  &13.415  & RRab$^*$\\
BW10 V27   & 0.25828  &15.601  &  1.230  &13.757  & RRc\\
BW10 V36   & 0.27700  &15.667  &  1.280  &13.747  & RRc\\
BW10 V39   & 0.51363  &16.018  &  1.646  &13.549  & RRab$^*$\\
BW10 V40   & 0.68277  &15.757  &  1.385  &13.679  & RRab$^*$\\
BW10 V41   & 0.62690  &15.972  &         &        & RRab$^*$\\
BW10 V44   & 0.52727  &15.977  &  1.501  &13.725  & RRab$^*$\\
BW10 V45   & 0.25283  &15.826  &  1.189  &14.043  & RRc\\
BW10 V46   & 0.32542  &15.924  &  1.370  &13.868  & RRc\\
BW10 V48   & 0.32333  &15.910  &  1.486  &13.681  & RRc\\
BW10 V51   & 0.27651  &15.965  &  1.262  &14.071  & RRc\\
BW10 V56   & 0.61826  &15.971  &  1.542  &13.658  & RRab$^*$\\
BW10 V59   & 0.31930  &16.112  &  1.237  &14.257  & RRc\\
BW10 V60   & 0.30371  &16.158  &  1.544  &13.842  & RRc\\
BW10 V65   & 0.28099  &16.190  &  1.073  &14.580  & RRc\\
BW10 V66   & 0.58123  &16.351  &  1.484  &14.125  & RRab$^*$\\
BW10 V70   & 0.28301  &16.216  &  1.353  &14.186  & RRc\\
BW10 V95   & 0.63810  &16.664  &  1.603  &14.261  & RRab$^*$\\
\hline
BW11 V3    & 0.59598  &14.590  &  1.284  &12.664  & RRab$^*$\\
BW11 V10   & 0.45066  &15.656  &  1.522  &13.373  & RRab$^*$\\
BW11 V17   & 0.49491  &15.852  &  1.357  &13.817  & RRab$^*$\\
BW11 V23   & 0.27402  &15.632  &  1.255  &13.750  & RRc\\
BW11 V25   & 0.51660  &15.969  &  1.385  &13.892  & RRab$^*$\\
BW11 V29   & 0.46610  &16.055  &  1.580  &13.685  & RRab$^*$\\
BW11 V34   & 0.34167  &15.834  &  1.386  &13.755  & RRc\\
BW11 V36   & 0.51394  &16.124  &  1.552  &13.796  & RRab$^*$\\
BW11 V38   & 0.59301  &15.915  &  1.488  &13.683  & RRab$^*$\\
BW11 V39   & 0.31226  &15.971  &  1.232  &14.123  & RRc\\
BW11 V44   & 0.48248  &16.259  &  1.521  &13.978  & RRab$^*$\\
BW11 V55   & 0.28618  &16.243  &  1.459  &14.053  & RRc\\
\hline
\end{tabular}
}
\end{center}
\pagebreak

\begin{center}
Table 9 \\
 
Parameters of RR Lyr stars from MM5 and MM7 fields. \\
\vspace{7pt}
 
{\scriptsize 
\begin{tabular}{|l|c|c|c|c|c|}
\hline
Star & Period & $\langle I \rangle$ & $\langle V-I \rangle$ & $V_{V-I}$
 & Type \\
\hline

 MM5-A V6   & 0.51202  & 15.491  & 1.781 & 12.819 & RRab$^*$ \\
 MM5-A V9   & 0.51619  & 15.696  & 1.368 & 13.643 & RRab$^*$ \\
 MM5-A V15  & 0.60022  & 15.720  & 1.282 & 13.798 & RRab$^*$ \\
 MM5-A V18  & 0.52667  & 15.969  & 1.597 & 13.573 & RRab$^*$ \\
 MM5-A V19  & 0.61251  & 15.929  & 1.466 & 13.730 & RRab$^*$ \\
 MM5-A V20  & 0.39119  & 15.706  & 1.171 & 13.949 & RRc \\
 MM5-A V21  & 0.46382  & 16.119  & 1.589 & 13.735 & RRab$^*$ \\
 MM5-A V27  & 0.62157  & 16.011  & 1.468 & 13.809 & RRab$^*$ \\
 MM5-A V32  & 0.54469  & 16.307  & 1.514 & 14.035 & RRab$^*$ \\
 MM5-A V37  & 0.45791  & 16.616  & 1.483 & 14.391 & RRab$^*$ \\
 MM5-A V41  & 0.58450  & 16.546  & 1.196 & 14.751 & RRab$^*$ \\
 MM5-A V46  & 0.37019  & 16.524  & 1.316 & 14.549 & RRab$^*$ \\
\hline
 MM5-B V4   & 0.52489  & 15.202  & 1.245 & 13.333 & RRab$^*$ \\
 MM5-B V10  & 0.52391  & 15.814  & 1.287 & 13.884 & RRab$^*$ \\
 MM5-B V20  & 0.64994  & 15.878  & 1.326 & 13.889 & RRab$^*$ \\
 MM5-B V21  & 0.50359  & 16.096  &   --   &   --    & RRab \\
 MM5-B V28  & 0.47995  & 16.266  & 0.932 & 14.868 & RRab$^*$ \\
 MM5-B V31  & 0.36456  & 16.589  & 1.041 & 15.027 & RRab \\
 MM5-B V41  & 0.25745  & 16.299  & 1.035 & 14.747 & RRc \\
 MM5-B V44  & 0.52311  & 16.667  & 1.205 & 14.860 & RRab$^*$ \\
\hline
 MM7-A V2  & 0.54078  & 15.141  & 1.153 & 13.411 & RRab$^*$ \\
 MM7-A V20 & 0.52984  & 16.036  & 1.561 & 13.694 & RRab$^*$ \\
 MM7-A V25 & 0.49380  & 15.965  &   --   &   --    & RRab \\
 MM7-A V26 & 0.44361  & 16.166  & 1.454 & 13.985 & RRab$^*$ \\
 MM7-A V34 & 0.53049  & 16.052  & 1.414 & 13.931 & RRab$^*$ \\
 MM7-A V44 & 0.22759  & 16.127  & 1.056 & 14.543 & RRc \\
 MM7-A V47 & 0.28782  & 16.200  & 1.291 & 14.264 & RRc \\
 MM7-A V94 & 0.27499  & 17.041  & 1.361 & 14.999 & RRc \\
\hline
 MM7-B V6  & 0.53913  & 15.670  & 1.367 & 13.619 & RRab$^*$ \\
 MM7-B V7  & 0.59002  & 15.733  & 1.312 & 13.765 & RRab$^*$ \\
 MM7-B V8  & 0.55940  & 15.694  & 1.362 & 13.650 & RRab$^*$ \\
 MM7-B V12 & 0.88121  & 15.698  & 1.500 & 13.447 & RRab$^*$ \\
 MM7-B V13 & 0.53321  & 15.948  & 1.378 & 13.881 & RRab$^*$ \\
 MM7-B V14 & 0.28675  & 15.764  & 1.212 & 13.947 & RRc \\
 MM7-B V63 & 0.61806  & 17.387  & 1.539 & 15.078 & RRab \\
\hline
\end{tabular}
}
\end{center}
\pagebreak

The map of interstellar extinction for BWC--BW8 OGLE fields given by Stanek 
(1996) was used to calculate free from interstellar extinction values of $I_0$ 
and ${(V-I)_0}$. Such data are presented in Tables~7--9. Each Table contains 
star designation, its period $P$ in days, ${\langle I\rangle}$ (magnitude 
integrated) brightness, ${V-I}$ color (defined as ${\langle V\rangle-\langle 
I\rangle}$), $I_0$ brightness, ${(V-I)_0}$ color (last two values are 
available only for BWC--BW8 fields) and $V_{V-I}$ parameter (see Section~3). 
The $V-I$ colors presented in this paper are slightly different than those 
given in the Catalog. This is a result of using in this work new $V$-band data 
collected in 1995 (not included in the first three parts of the Catalog) and 
of computing the mean value of ${V-I}$ instead of ${V-I}$ color at maximum 
brightness. 

For a few stars ${V-I}$ color is not determined because of lack of the 
accurate \hbox{$V$-band} photometry. These stars are excluded from further analysis. 
We also omitted faint RR~Lyr variables which are located behind 
the Galactic bulge (i.e. BW4~V132, BW5~V135, BW5~V174, MM7-B~V63). Some of them 
most likely belong the to Sagittarius Dwarf Galaxy (see Alard 1996). 
Describing properties of RR~Lyr variables we used mainly type ab of these 
stars. Clear discrimination between both types of RR~Lyr stars, based only on 
amplitude defined as combination of Fourier coefficients (${{\rm amp}=\sqrt 
{a_1^2+b_1^2}}$) and period of the star, is presented in Fig.~2. Recently 
Minniti {\it et al.} (1996) published preliminary results concerning RR~Lyr variables 
in MACHO Collaboration database. They found three peaks in period distribution 
of RR~Lyr stars in Large Magellanic Cloud (LMC) and Galactic bulge. The 
highest peak with longest value of period corresponded to RRab stars 
(variables pulsating in fundamental mode), the second peak near period 
0.32--0.34$^d$ corresponded to RRc stars (pulsating in 1st 
overtone) and the smallest peak with period around 0.27--0.28$^d$
was interpreted as RRe stars (pulsations in 2nd overtone). 

Similar distribution for OGLE RR~Lyr stars from Galactic bulge is 
presented in Fig.~3. One can clearly distinguish two peaks: the higher 
corresponding to RRab stars and smaller corresponding to RRc variables. No 
clear peak connected with RRe stars is detected but assymetry 
in the shape of RRc peak might be  caused by RRe stars. 
 
All stars from Tables~7--9 used in further calculations are marked by asterisk. 
The objects from Tables~7--9 are presented in color magnitude diagram shown in 
Fig.~4. For comparison 20\% of non-variable stars from BWC field are also 
plotted. 

\vspace*{-7pt}
\section{$V_{V-I}$ Parameter for RR~Lyr Stars in the Galactic Bulge}
\vspace*{-7pt}

Stanek {\it et al.} (1994) used a well-defined population of bulge red clump stars to 
investigate the inner galactic structure. For this purpose they defined the 
extinction-insensitive $V_{V-I}$ parameter: 
\vspace*{-3pt}
$$V_{V-I}\equiv V-2.6\cdot (V-I)\eqno(2)$$
with reddening law: ${E_{V-I}=A_V/2.6}$. The values of $V_{V-I}$ 
showed clear correlation with galactic longitude $l$, where the smallest 
$V_{V-I}$ were found in fields MM7 (${l\approx5.5^\circ}$) and highest ones 
in fields MM5 (${l\approx-4.8^\circ}$). The difference in values of $V_{V-
I}$ between those fields amounted to 0.4~mag which implied the presence of a 
bar with the major axis inclined to the line of sight by 20--30~deg with axis 
ratios 3.5:1.5:1 (Stanek {\it et al.} 1997). In their discussion Stanek {\it et al.} (1994) 
pointed out that the changes of the brightness and color of red clump stars, 
which suggest presence of the bar, might be also caused by variations of 
$A_V/E_{V-I}$ ratio. In the case when this ratio in BW fields is equal to 2.6, the 
values ${\sim2.9}$ in MM5 fields and ${\sim2.3}$ for MM7 fields would explain 
behavior of $V_{V-I}$ parameter without assumption of presence of the Galactic 
bar. 

Another group of stars which might be used to measure $V_{V-I}$ in different 
fields of the Galactic center are RR~Lyr stars. Using our sample and 
according to a new determination of ${A_V/E_{V-I}}$ ratio made by 
Wo\'zniak and Stanek (1996) and Stanek (1996) we took: 
\vspace*{-3pt}
$$V_{V-I}\equiv V-2.5\cdot(V-I)\eqno(3)$$
and computed the values of $V_{V-I}$ presented in Tables~7--9.

Due to small number of RR~Lyr stars in comparison to red clump giants we 
decided to calculate mean and median values of $V_{V-I}$. In calculations 
we used only ab type of RR Lyr stars marked in Tables~7--9 by asterisk. Results are 
presented in Fig.~5. No clear correlation between $V_{V-I}$ and galactic 
longitude $l$ in both cases is detected. It seems to suggest that there is no 
fluctuations in $A_V/E_{V-I}$ ratio, the bar really exists and RR~Lyr 
variables form a spherically symmetric inner halo around center of our Galaxy. 
The rims of that inner halo are most likely a beginning of the outer halo 
containing also RR~Lyr variables and additionally globular clusters and blue 
horizontal branch stars. But accuracy of our determination of $V_{V-I}$ for RR~Lyr 
stars is low due to limited sample and one cannot totally exclude possibility 
of the presence of some dependence $V_{V-I}$ on $l$ for RR~Lyr stars. Similar 
result for their $W_{V-R}$ extinction insensitive parameter for RR~Lyr stars 
was obtained by Minniti {\it et al.} (1996).

\section{Conclusions}

We presented parameters of the Fourier decomposition of light curves of RR~Lyr 
type variable stars from The OGLE Catalog of Periodic Variable Stars in the Galactic 
Bulge. We also calculated  values of mean ${\langle I\rangle}$ brightness, ${V-I}$ 
color and extinction insensitive $V_{V-I}$ parameter for all RR~Lyr stars in 
the Catalog. Additionally for fields with known interstellar extinction (Stanek 
1996) i.e. BWC--BW8 we computed $I_0$ and ${(V-I)_0}$. Based on Fourier 
parameters we showed the way of clear discrimination between RRab and RRc 
stars. The construction of histogram with period distribution allowed us to 
find trace of third kind of RR~Lyr stars (RRe pulsating in the 2nd overtone). 
Similar result was reported by Minniti {\it et al.} (1996) in population of RR~Lyr 
stars in LMC and the Galactic bulge. 

To check the stability of the $A_V/E_{V-I}$ ratio in the 
fields in the Galactic bulge we also investigated the behavior of extinction 
insensitive parameter $V_{V-I}$ for RR~Lyr variables placed at different 
galactic longitudes. There is no clear trend between $V_{V-I}$ and galactic 
longitude $l$ but due to the small number of objects in our sample accuracy of 
determined mean values of $V_{V-I}$ is low. If the lack of $V_{V-I}(l)$ 
dependence is real it means that RR~Lyr variables form a spherically symmetric 
inner halo around center of our Galaxy. It also means that $A_V/E_{V-I}$ ratio 
in all measured OGLE fields is stable and likely equal to 2.5 (Wo\'zniak and 
Stanek 1996). This fact confirms the presence of the galactic bar which was 
found in the investigation of the red clump stars (Stanek {\it et al.} 1994, 1997). 

To confirm above conclusions further observations of RR~Lyr stars, especially 
in the fields with ${|l|>5^\circ}$, are needed. 

All photometric parameters of RR~Lyr stars presented in Tables~2--6 and 7--9 
are available in electronic form via INTERNET from the following URL: 
\linebreak{\it http://www.astrouw.edu.pl/$\sim$olech/rrlyr.html}. 

\vspace{0.4cm}
\noindent {\bf Acknowledgments} It is a great pleasure to thank Prof. Bohdan Paczy\'nski and Dr 
Andrzej Udalski for their continuous and helpful discussions, reading and 
commenting on the manuscript. The paper was partly supported by the KBN BW 
grant to the Warsaw University Observatory.
 
\vspace{0.7cm}
\begin{center}
REFERENCES
\end{center}
\vspace{0.5cm}

\noindent{Alard, C.},~{1996},~{ApJL},~{458},~{17}

\noindent{Minniti {\it et al.}},~{1996},~{astro-ph/9610025}

\noindent{Olech, A.},~{1996},~{Acta Astron.},~{46},~{389}

\noindent{Paczy\'nski, B.},~{1986},~{ApJ},~{304},~{1}

\noindent{Paczy\'nski, B., and Udalski, A.},~{1996},~{The proceedings
of 12th IAP Colloquium: Variable Stars and the Astrophysical Returns of
Microlensing Surveys}

\noindent{Ruci\'nski, S.},~{1997a},~{AJ},~{113},~{407}

\noindent{Ruci\'nski, S.},~{1997b},~{,~},~{,~},~{astro-ph/9611158}

\noindent{Stanek, K.Z.},~{1996},~{ApJL},~{460},~{37}

\noindent{Stanek, K.Z., Mateo, M., Udalski, A., Szyma\'nski, M.,
Ka{\l}u\.zny, J., Kubiak.,M.},~{1994},~{ApJL},~{429},~{73}

\noindent{Stanek, K.Z., Udalski, A., Szyma\'nski, M.,
Ka{\l}u\.zny, J., Kubiak., Mateo, M., Krzemi\'nski, W.},~{1997},~{ApJ},~{477},~{163}

\noindent{Szyma\'nski, M., Udalski, A., Kubiak, M., Ka{\l}u\.zny, J.,
Mateo, M., and Krzemi\'nski, W.},~{1996},~{Acta Astron.},~{46},~{1}

\noindent{Udalski, A., Szyma\'nski, M., Ka{\l}u\.zny, J., Kubiak, M., and
Mateo, M.},~{1992},~{Acta Astron.},~{42},~{253}

\noindent{Udalski, A., Szyma\'nski, M., Ka{\l}u\.zny, J., Kubiak, M., and
Mateo, M.},~{1993a},~{Acta Astron.},~{43},~{69}

\noindent{Udalski, A., Szyma\'nski, M., Ka{\l}u\.zny, J., Kubiak, M.,
Krzemi\'nski, W., Mateo, M., Preston, G.W., and Paczy\'nski,
B.},~{1993b},~{Acta Astron.},~{43},~{289}

\noindent{Udalski, A., Szyma\'nski, M., Stanek, K.Z., Ka{\l}u\.zny, J.,
Kubiak, M., Mateo, M., Krzemi\'nski, W., Paczy\'nski, B., and Venkat,
R.},~{1994a},~{Acta Astron.},~{44},~{165}

\noindent{Udalski, A., Szyma\'nski, M., Ka{\l}u\.zny, J., Kubiak, M.,
Mateo, M., Krzemi\'nski, W., and Paczy\'nski, B.},~{1994b},~{Acta Astron.},~{44},~{227}

\noindent{Udalski, A., Kubiak, M., Szyma\'nski, M., Ka{\l}u\.zny, J.,
Mateo, M., and Krzemi\'nski, W.},~{1994c},~{Acta Astron.},~{44},~{317}

\noindent{Udalski, A., Szyma\'nski, M., Ka{\l}u\.zny, J., Kubiak, M.,
Mateo, M., and Krzemi\'nski, W.},~{1995a},~{Acta Astron.},~{45},~{1}

\noindent{Udalski, A., Olech, A., Szyma\'nski, M., Ka{\l}u\.zny, J.,
Kubiak, M., Mateo, M., and Krzemi\'nski, W.},~{1995b},~{Acta Astron.},~{45},~{433}

\noindent{Udalski, A., Olech, A., Szyma\'nski, M., Ka{\l}u\.zny, J.,
Kubiak, M., Mateo, M., Krzemi\'nski, W., and Stanek K.Z.},~{1996},~{Acta Astron.},~{46},~{51}

\noindent{Udalski, A., Olech, A., Szyma\'nski, M., Ka{\l}u\.zny, J.,
Kubiak, M., Mateo, M., Krzemi\'nski, W., and Stanek K.Z.},~{1997},~{Acta Astron.},~{47},~{1}

\noindent{Wo\'zniak, P., and Stanek, K.Z.},~{1996},~{ApJ},~{464},~{233}

\pagebreak
\begin{center}
FIGURE CAPTION
\end{center}
\vspace{0.5cm}
 
\begin{itemize}

\item[Fig. 1.] Light curves of six RR~Lyr stars from OGLE BW1 field.
        Solid line corresponds to the fit given by Eq.~(1).

\item[Fig. 2.] Correlation between period and amplitude for RR~Lyr stars
      from the Galactic bulge with clear discrimination between RRab and RRc variables.
      Open circles correspond to RRc variables and filled circles to RRab stars.

\item[Fig. 3.] The period distribution for RR~Lyr stars in the Galactic bulge.

\item[Fig. 4.] Color--magnitude diagram for RR~Lyr stars. Stars marked
       by open circles are located in BW fields. Crosses and open triangles correspond to
       RR~Lyr variables from MM5 and MM7 fields, respectively.

\item[Fig. 5.] Correlation between extinction insensitive parameter
       $V_{V-I}$ and galactic longitude $l$ for RRab stars in OGLE database. See for
       comparison Minniti et al. (1996).
\end{itemize}

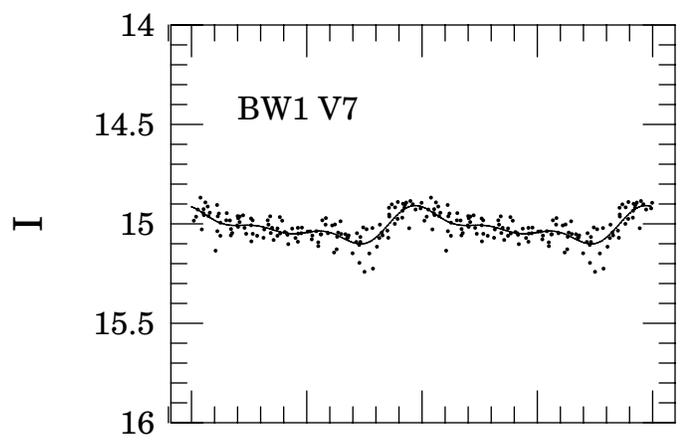
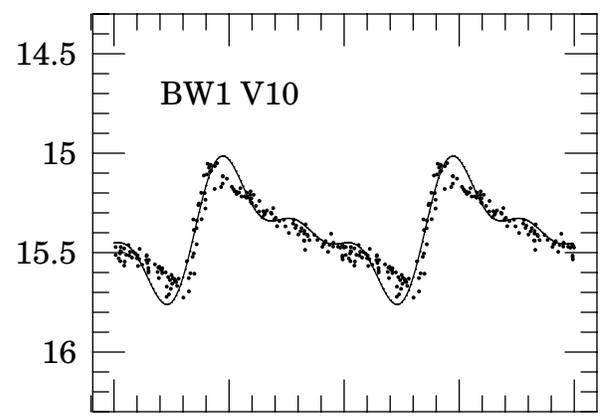
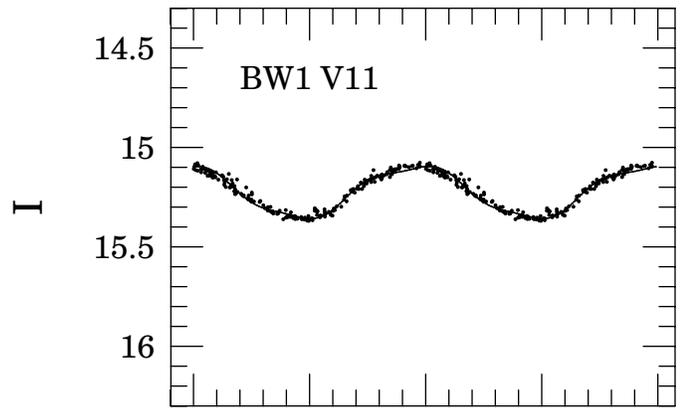
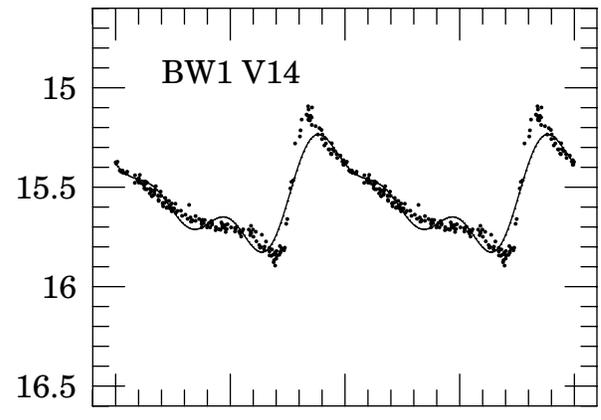
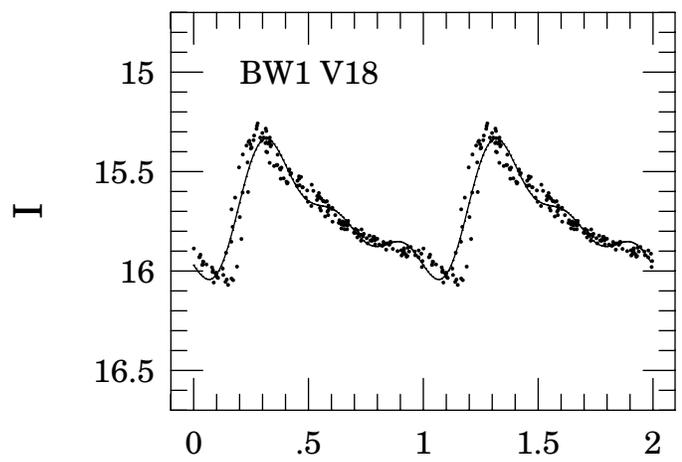
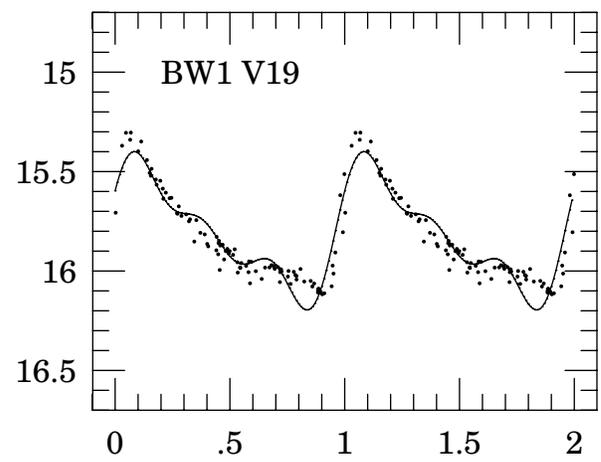
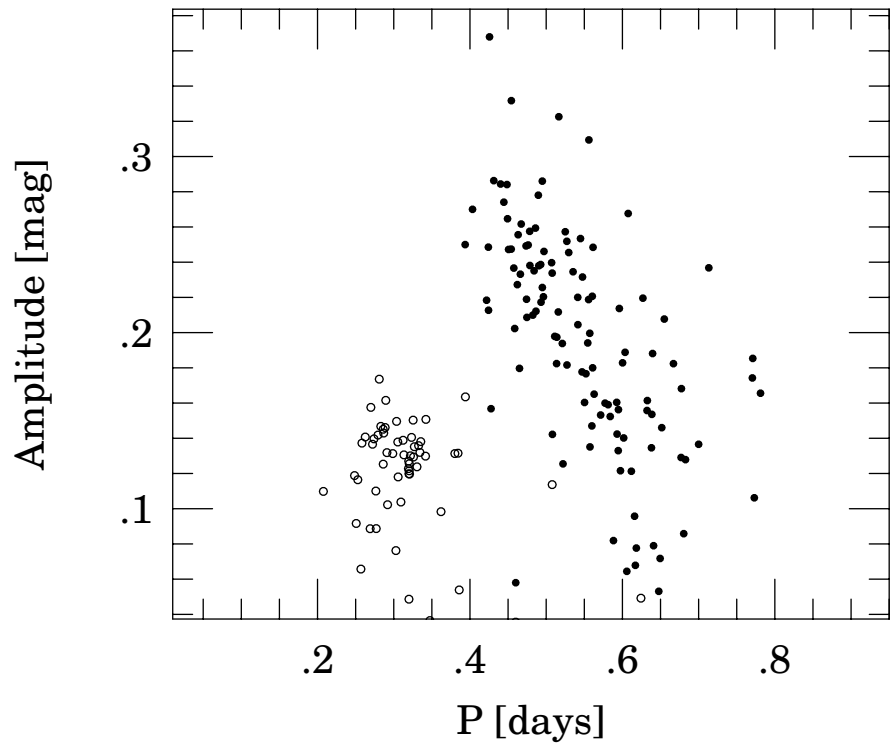
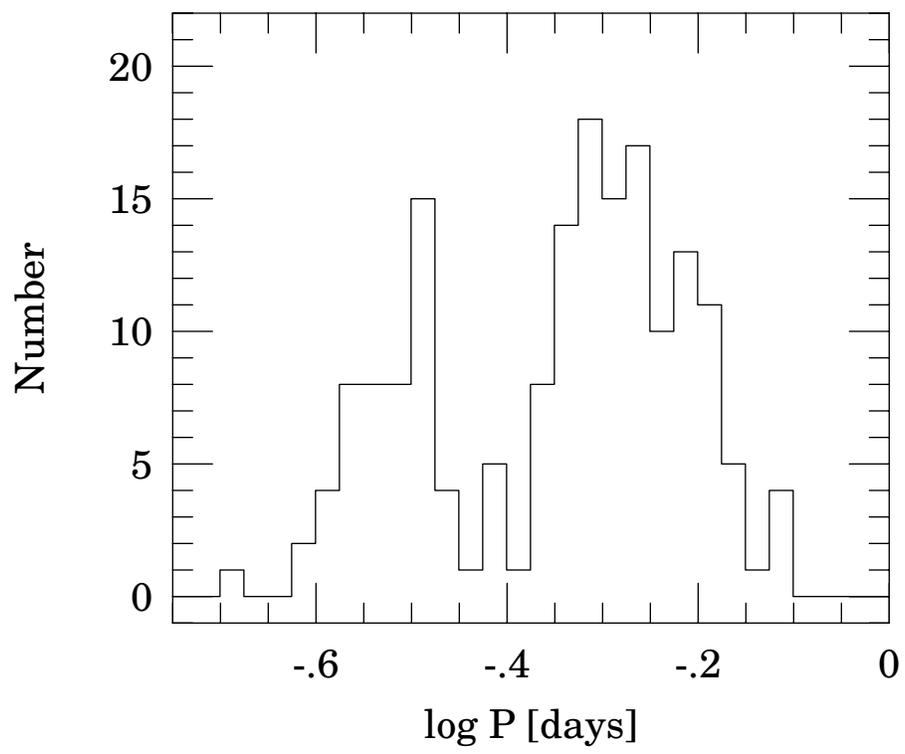
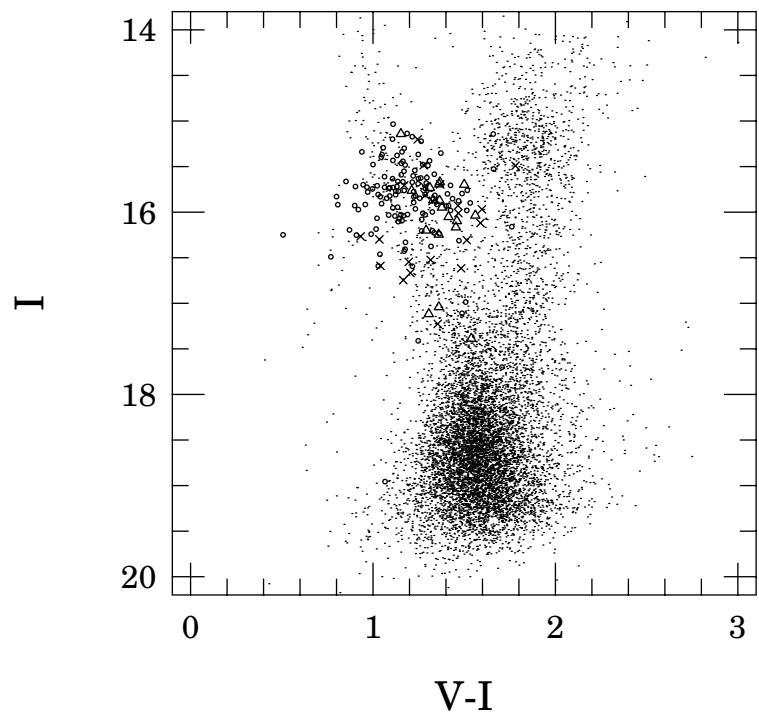
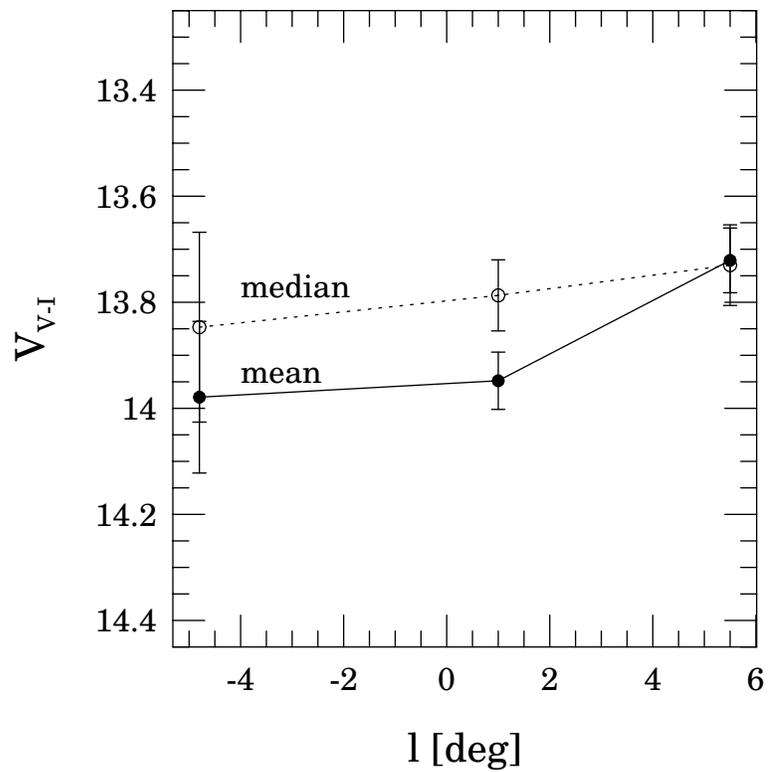
\end{document}